\documentclass[psfig,epsf,rotate]{aa}
\usepackage{graphicx}
\begin{document}
\thesaurus{08 (08.16.5;                     
            09.01.1;                     
            09.09.1 Orion IRc2;          
            09.10.1;                     
            09.13.2;                     
            13.09.4)}                    
 
 
 
\title{ISO--SWS observations of pure rotational H$_2$O absorption lines
toward Orion--IRc2 \thanks{Based on observations with ISO, an ESA project with
instruments funded by ESA Member States (especially the PI countries: France, 
Germany, The Netherlands and the United Kingdom) and with the participation of 
ISAS and NASA.}}
 
\author{C.M. Wright\inst{1,2}
\and E.F. van Dishoeck\inst{1} 
\and J.H. Black\inst{3}
\and H. Feuchtgruber\inst{4,5}
\and J. Cernicharo\inst{6}
\and E. Gonz\'alez-Alfonso\inst{6,7}
\and Th. de Graauw\inst{8}}

\offprints{C.M. Wright (wright@ph.adfa.edu.au)}
 
\institute{Leiden Observatory, P.O.\ Box 9513, 2300 RA Leiden, The Netherlands
\and School of Physics, University College, Australian Defence
Force Academy, University of New South Wales, Canberra ACT 2600, Australia
\and Onsala Space Observatory, Chalmers University of Technology, S--43992 
Onsala, Sweden
\and ISO--SOC, ESA Astrophysics Division, P.O.\ Box 50727, E--28080 
Villafranca/Madrid, Spain
\and Max--Planck--Institut f\"{u}r extraterrestrische Physik,
Postfach 1603, 85740 Garching bei M\"{u}nchen, Germany
\and CSIC--IEM, Serrano 121, E--28006 Madrid, Spain
\and Universidad de Alcal\'a de Henares, Dept.\ de F{\'{\i}}sica,
Campus Universitario, E--28871 Alcal\'a de Henares, Madrid, Spain
\and SRON, P.O.\ Box 800, NL--9700 AV Groningen, The Netherlands}
\date{Received date: accepted date}
\titlerunning{Water absorption toward Orion IRc2}
\maketitle



\begin{abstract}
 
First detections of thermal water vapor absorption lines have been made 
toward Orion IRc2 using the {\it Short Wavelength Spectrometer} (SWS) on 
board the {\it Infrared Space Observatory} (ISO). Grating spectra 
covering wavelengths 25--45 $\mu$m yield 19 pure rotational lines, 
originating from energy levels 200--750~K above ground. Fabry-Perot 
spectra of 5 transitions resolve the line profiles and reveal the H$_{2}$O 
gas kinematics. The fact that all lines are seen in absorption 
is in striking contrast with data from the ISO {\it Long Wavelength 
Spectrometer} (LWS), where the H$_2$O lines appear in emission. At 
least one line displays a P-Cygni type profile, which suggests that the 
water is located in an expanding shell centered on or near IRc2. The expansion 
velocity is 18 km s$^{-1}$, in agreement with the value inferred from 
H$_{2}$O maser observations by Genzel et al.\ (1981). Because the continuum 
is intense and likely formed in or near the water-containing gas, the 
excitation of the observed transitions is dominated by radiative processes. 
A simple, generalised curve-of-growth method is presented and used to 
analyze the data. A mean excitation temperature of 72 K and a total H$_2$O 
column density of $1.5\times 10^{18}$ cm$^{-2}$ are inferred, each with 
an estimated maximum uncertainty of 20\%. Combined with the H$_2$ column 
density derived from ISO observations of the pure rotational H$_2$ lines, 
and an assumed temperature of 200--350 K, the inferred H$_2$O 
abundance is 2--5$\times 10^{-4}$ in the warm shocked gas. This abundance 
is similar to that found recently by Harwit et al. (1998) toward Orion 
using data from the LWS, but higher than that found for most other shocked 
regions by, for example, Liseau et al. (1996).

 
\end{abstract}
  
\section{Introduction}
 
Water is one of the prime species for probing the interaction between
young stars and their surroundings, both in terms of its abundance
and its peculiar excitation. In high temperature gas, appropriate to
shocks and hot cores for instance, all of the oxygen not locked up in
CO is predicted to be driven into H$_2$O at temperatures above 230 K by the
O + H$_2$ $\to$ OH + H and OH + H$_2$ $\to$ H$_2$O + H reactions, 
resulting in very bright H$_2$O lines (e.g., Hollenbach \& McKee 1979, 
Neufeld \& Melnick 1987, Kaufman \& Neufeld 1996, Charnley 1997).
In contrast, the H$_2$O abundance may be at least two orders of magnitude 
lower in surrounding colder gas (e.g., Zmuidzinas et al.\ 1995).
In addition to collisions, the H$_2$O excitation and line profiles can be 
strongly affected by mid- and far-infrared radiation from warm dust 
(e.g., Phillips et al.\ 1978, Takahashi et al. 1983, 1985),
providing detailed information on the physical parameters of the gas
and its location with respect to the radiation sources.

Interstellar water has been difficult to detect, apart from its presence as an
ice mantle on dust grains or through its maser emission, owing to the severe
telluric absorption encountered at Earth-based observatories. Nevertheless,
over the last twenty years, many searches for lines of gas-phase H$_2$O and
its isotopomers have been made from the ground and airborne altitudes, in
particular toward Orion (e.g., Waters et al.\ 1980, Phillips et al.\ 1978,
Jacq et al. 1990, Wannier et al.\ 1991, Cernicharo et al.\ 1994, Zmuidzinas et
al.\ 1995, Tauber et al.\ 1996, Timmermann et al.\ 1996, Gensheimer et al.\
1996, Cernicharo et al.\ 1999a,b). However, due to the choice of line, 
wavelength and beam size, different components are often probed, whilst many
ground-based observations refer to masing lines, for which 
sophisticated shock and maser models are required in order to extract 
physical parameters. 
 
One of the major aims of the {\it Infrared Space Observatory} ({\it ISO}) 
mission has been the routine measurement of thermal gas-phase water
lines, and their use as diagnostics of the chemical and physical conditions 
within molecular clouds. Far-infrared pure rotational H$_2$O emission lines 
in the 50--200 $\mu$m wavelength range have been detected with the {\it Long
Wavelength Spectrometer} (LWS) of ISO in a number of star-forming regions 
(e.g., Liseau et al.\ 1996, Ceccarelli et al.\ 1998), including Sgr B2 
(Cernicharo et al.\ 1997a) and Orion (Cernicharo et al.\ 1997b, 1999a; 
Harwit et al.\ 1998). Additionally, van Dishoeck \& Helmich (1996), van 
Dishoeck (1998) and Dartois et al.\ (1998) have observed absorption around 
6 $\mu$m in the $\nu_{2}$=1--0 band toward a number of deeply embedded, 
massive young stars. Typical H$_2$O abundances of $10^{-5}$ with respect 
to H$_2$ have been derived from these data.
Similar observations have recently been reported for Orion BN/IRc2
by van Dishoeck et al.\ (1998) and Gonz\'{a}lez-Alfonso et al.\ (1998),
although in this case emission is also detected. Wright et al.\ (1997) have
however shown that the detection of the corresponding pure rotational lines
at $\sim$ 30--200 $\mu$m in most sources observed at 6 $\mu$m is still
difficult. The observations of Orion-IRc2 presented here form a
notable exception.
  
Many of the earlier searches for H$_2$O lines have been performed toward the
BN/IRc2 complex of infrared sources in Orion, because of the extraordinary
brightness of many atomic and molecular lines in this region compared with
other clouds (e.g., Genzel \& Stutzki 1989, Blake 1997). See van Dishoeck et
al.\ (1998) and references therein for a detailed description of the geometry
and the diverse range of physical phenomena in this region. In this paper we
present the first detection of numerous pure rotational water lines in
absorption toward IRc2 in the 25--45 $\mu$m interval with the {\it Short
Wavelength Spectrometer} (SWS) (de Graauw et al.\ 1996) on board {\it ISO}.
Some of these lines have been velocity resolved with the Fabry-Perot,
enabling direct information on the location of the absorbing gas to be 
inferred. These data complement the earlier ground-based data, as well as 
observations of pure rotational lines with the LWS obtained by Cernicharo 
et al.\ (1997b, 1999a) and Harwit et al.\ (1998) in a much larger beam.

Our interpretation of the ISO spectra of Orion IRc2 suggests that the 25--45
$\mu$m H$_2$O spectrum originates in a region where the intrinsically strong 
lines couple efficiently to an intense continuum. In principle, the formation 
of such a spectrum should be described for a stratified atmosphere in which 
lines and continuum are treated self-consistently. We show here that the 
observed features of the spectrum can be described well by a simple 
``generalised curve-of-growth'', which includes the effects of coupling to 
a strong continuum. In the following we describe our observations and 
present our results, followed by a discussion of the location of the 
absorbing water vapour, its excitation and finally its abundance.
 
\section{Observations and data reduction}
 
\subsection{Observations}

A complete grating scan from 2.4--45.3 $\mu$m using the maximum spectral
resolution SWS06 observing mode was made on September 6 1997 (revolution 660)
centered at
$\alpha(2000)=05^{h} 35^{m} 14.2^{s}$, $\delta(2000)=-05^{\circ} 22' 31.5''$,
which is about 3$''$ W and 1$''$ S of the IRc2 position quoted by Downes
et al.\ (1981). The long axis of the ISO--SWS aperture was oriented at
172.81$^{\circ}$ E of N. The aperture size varies from $14''\times 20''$ at
2.4--12 $\mu$m, to $14''\times27''$ at 12--27.5 $\mu$m, $20''\times27''$ at 
27.5--29 $\mu$m and $20''\times33''$ at 29--45.2 $\mu$m. A further SWS06 
scan was made from 26.3 to 45.2 $\mu$m in revolution 826 (February 19 1998) 
with the long axis oriented at 165.79$^{\circ}$ E of N. The SWS beam 
includes both IRc2 and BN, but not ``peak 1'' or ``peak 2'' of shocked 
H$_2$ (Beckwith et al.\ 1978).
 
The resolving power of the ISO--SWS grating, $\lambda/\Delta \lambda$, varies
from about 1000--2500, implying that the observed line profiles are not
resolved and that little velocity information can be obtained to help
disentangle the various components which are known to exist toward IRc2. 
For this reason, follow-up observations of a selection of H$_{2}$O
lines were obtained using the Fabry-Perot SWS07 observing mode in 
revolutions 823 and 831 (February 16 and 24 1998). Five 250 km s$^{-1}$ 
scans were made across each line. The resolving power is of order 30\,000, 
corresponding to a velocity resolution of 10 km s$^{-1}$, and the 
aperture long axis was oriented at 164.14$^{\circ}$ and 168.46$^{\circ}$ 
E of N on the two dates. The aperture size is $10''\times39''$ and 
$17''\times40''$ for wavelengths below and above 26 $\mu$m respectively. 
During the revolution 831 observation the wavelength interval between 5.3 
and 7.0 $\mu$m was simultaneously scanned by the SWS grating. Within this 
region are several H$_2$ lines (0--0 S(7), S(6) and S(5)), and H$_2$O solid 
state ice and H$_2$O gas phase $\nu_{2}$=1--0 features. 

\subsection{Data reduction}
 
Data reduction was carried out on the Standard Processed Data file 
from the Off Line Processing system (OLPv6.1.1 for revolution 660 and 
OLPv6.3.2 for revolutions 823, 826 and 831), using standard routines 
within the SWS Interactive Analysis package up to the Auto Analysis
Result (AAR) stage. The dark current subtraction was performed interactively
for the grating data, but not for the FP data. The most up-to-date 
calibration files available for wavelength, flux and relative spectral 
responsivity were used. Even so, there is at least one instrumental
artefact in our grating spectra, at 33 $\mu$m, due to structure in the 
relative spectral response calibration file (RSRF). 

The high continuum flux encountered in band 4 of the SWS grating
(29--45 $\mu$m), and the finite response time of the detectors, led to
significant memory and response effects near the band edges, at positions
of important water and OH lines, and so necessitated a non-standard
up-down scan correction. For the SWS an ``up'' (``down'') scan
proceeds toward increasing (decreasing) grating positions but
decreasing (increasing) wavelengths. Since the up scan precedes the
down scan, in high flux cases it is usually the down scan which is
deemed to have the more reliable spectral shape, since by that time
the instrument has settled down. Indeed, this was the case in the 
long-wavelength portion (41--45 $\mu$m) of our spectra, so that the up 
scan was corrected to approximately the same shape as the down scan. On 
the other hand, the short-wavelength portion (29--33 $\mu$m) of the down
scan of our spectra had a peculiar concave shape, which did not match
cleanly (in terms of {\em shape}) to band 3e, which in turn matched
cleanly with band 3d. 
This was true for both the revolution 660 and 826 data sets. In this case 
therefore the down scan was corrected to have a similar shape to the up 
scan. The up-down correction obviously entailed splitting the band 4 data 
into two segments (29--37 $\mu$m and 37--45 $\mu$m), but this produced a 
more reliable spectral shape at both ends, e.g.
allowing the continuum on either side of the 45.1 $\mu$m water line to 
be determined, and the 28.940 and 28.914 $\mu$m OH and H$_2$O $4_{40}-3_{13}$ 
lines to be partially resolved.

Following the AAR stage, further data processing, such as flatfielding, 
sigma clipping and rebinning, was performed using software in the SWS IA 
package. For instance, for both the grating and FP cases the individual 
detector scans have been brought to a common flux level using an $n^{\rm th}$ 
degree polynomial ($n$ = 0, 1, 2 or 3) fit to a reference spectrum which may
either be the median of the down scans (e.g. grating band 3), the median of 
both the up and down scans (e.g. the up-down corrected grating band 4), or 
the mean of all scans (e.g. the Fabry-Perot data). The data have
subsequently been sigma clipped such that any points lying more than 3
sigma outside of the average of all data within a bin of width equal
to the theoretical spectral resolution have been discarded. The final step 
in producing a spectrum involves rebinning the much over-sampled data using 
the mean within a bin which is again equal to the theoretical resolution. 

In the case of the 25.94 $\mu$m FP scans it was observed that glitches and 
their associated tails, which had not been flagged as bad data by the 
pipeline, contributed significant noise to the final spectrum. In this 
instance an interactive glitch and tail correction was performed before 
flatfielding, sigma clipping and rebinning. As noted by Schaeidt et al. 
(1996) and Heras (1997) the photometric accuracies of the SWS grating 
band 4 (where most of our water detections occur) and Fabry-Perot are 30\% 
and 40\% respectively. However, our observed line equivalent widths, being 
ratios of the line area over adjacent continuum, will be more accurate than 
this. The accuracy of the grating wavelength calibration is 1/10--1/5 of a 
resolution element, whilst that of the Fabry-Perot is of order 10$^{-4}$ 
$\mu$m, or about 1 km s$^{-1}$ (Valentijn et al. 1996; Feuchtgruber et al.
1997).

\section{Results}
 
In Fig. 1 the SWS grating spectrum in the 25--45 $\mu$m range is presented 
for the revolution 660 data, with the position and identification of 
features marked. The data consist of spectral segments from bands 3d, 3e 
and 4, for which ``jumps'' in the flux were observed due to the different 
aperture sizes. However, for display purposes small shifts, of order a few 
thousand Jy (i.e. a few percent), have been applied to produce a 
continuous spectrum. Clearly marked in Fig. 1 are 19 absorption lines 
arising from pure rotational levels in both the ortho- (o) and para- (p) 
species of H$_{2}$O. In the few cases where the plot scale makes the line 
difficult to see in Fig. 1, we present close-ups in Fig. 2. 
The lines in this wavelength region originate from highly-excited states, 
with excitation energies of the lower levels $hcE_{\ell}/k = 200-750$ K, 
and the upper levels up to 1200~K. The strongest absorptions come from 
levels belonging to the backbone of the energy ladder. Among the highest 
levels seen is the upper state of the $7_{25}-6_{16}$ transition at 
29.8367 $\mu$m, involving the upper level ($6_{16}$) of the well-known 
22 GHz H$_2$O maser line (Cheung et al.\ 1969). Indeed, many of the 
levels from which absorption is seen are also upper levels of either 
known or suspected maser transitions (Neufeld \& Melnick 1991). It is shown
below based on velocity considerations that our lines probably arise
in the same region where the masers are situated, indicating that our
observations may form a useful data set for studying the maser pump
mechanism. 

A sub-set of 5 of the 19 lines was chosen to be observed with the 
Fabry-Perot. These spectra are shown in Fig. 3. In one case
($4_{32}-3_{03}$ 40.69 $\mu$m) and possibly a second ($5_{41}-4_{32}$
43.89 $\mu$m), two velocity components are detected, one in absorption
and the other in emission. A modulation in the baselines, due to
the tracking of the Fabry-Perot transmission function across the
region of the peak of the grating instrumental profile, has been modelled
in terms of two parameters, the grating spectral resolution and the shift
between the two response functions. The spectra shown in Fig. 3 have been 
corrected for this tracking pattern. Also shown in
Fig. 3 is the grating spectrum from 5.3--7.0 $\mu$m obtained from
these new data. It is consistent with that already observed by van
Dishoeck et al. (1998) and Gonz\'{a}lez-Alfonso et al. (1998).

\begin{figure*}
\vspace*{0.25cm}
\hspace*{1.25cm}
\resizebox{14cm}{16cm}{\includegraphics{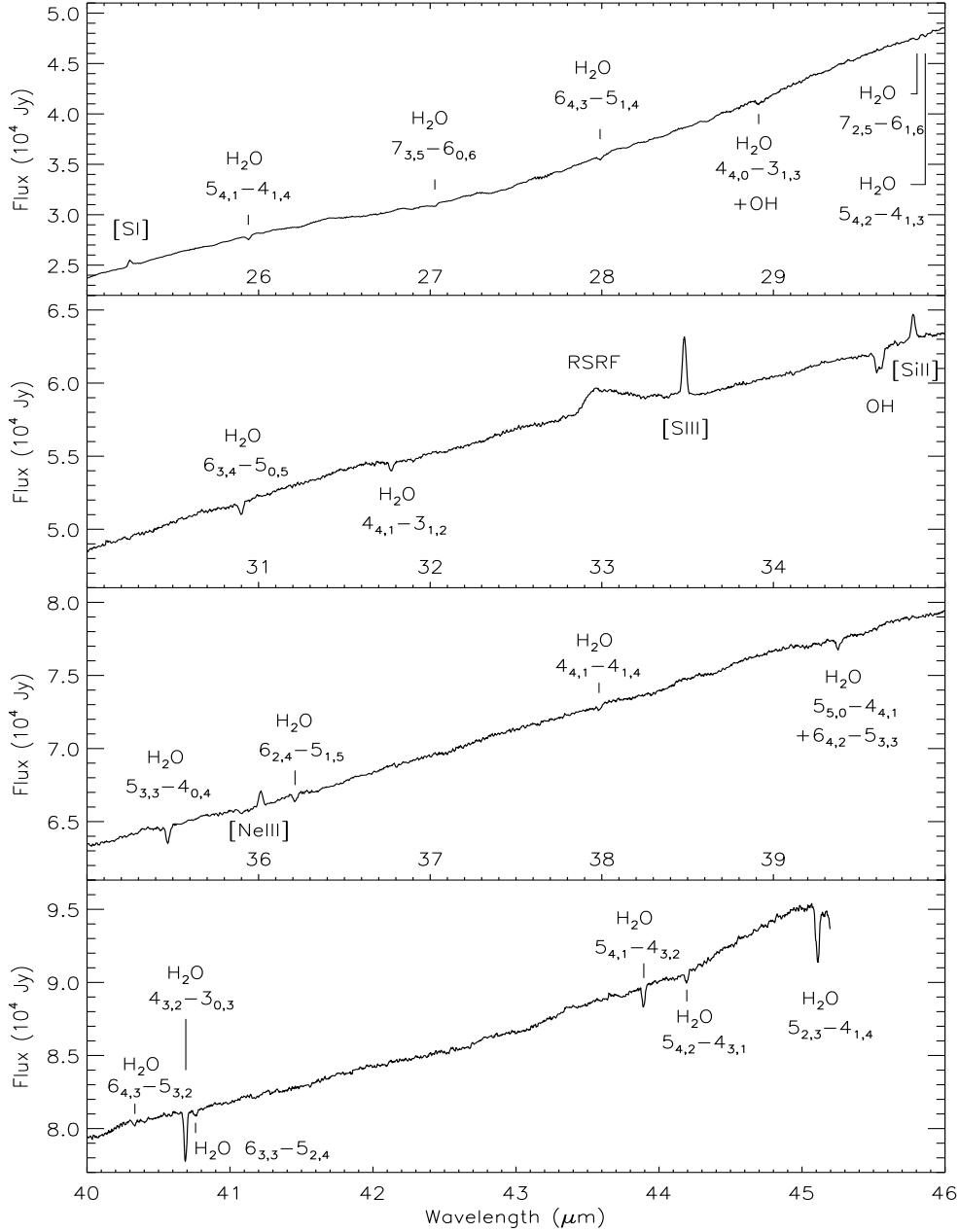}}
\vspace*{0.75cm}
\caption[]{ISO--SWS 25--45 $\mu$m grating spectrum toward Orion IRc2. The 
principal absorption and emission features are indicated. The data within 
bands 3d, 3e and 4 have been shifted so that they join continuously at the 
band edges. The feature marked RSRF is an instrumental artefact due to 
structure in the relative spectral response calibration file.}
\end{figure*}

\begin{figure*}
\vspace*{-0.75cm}
\hspace*{2.00cm}
\resizebox{18cm}{!}{\includegraphics[angle=90]{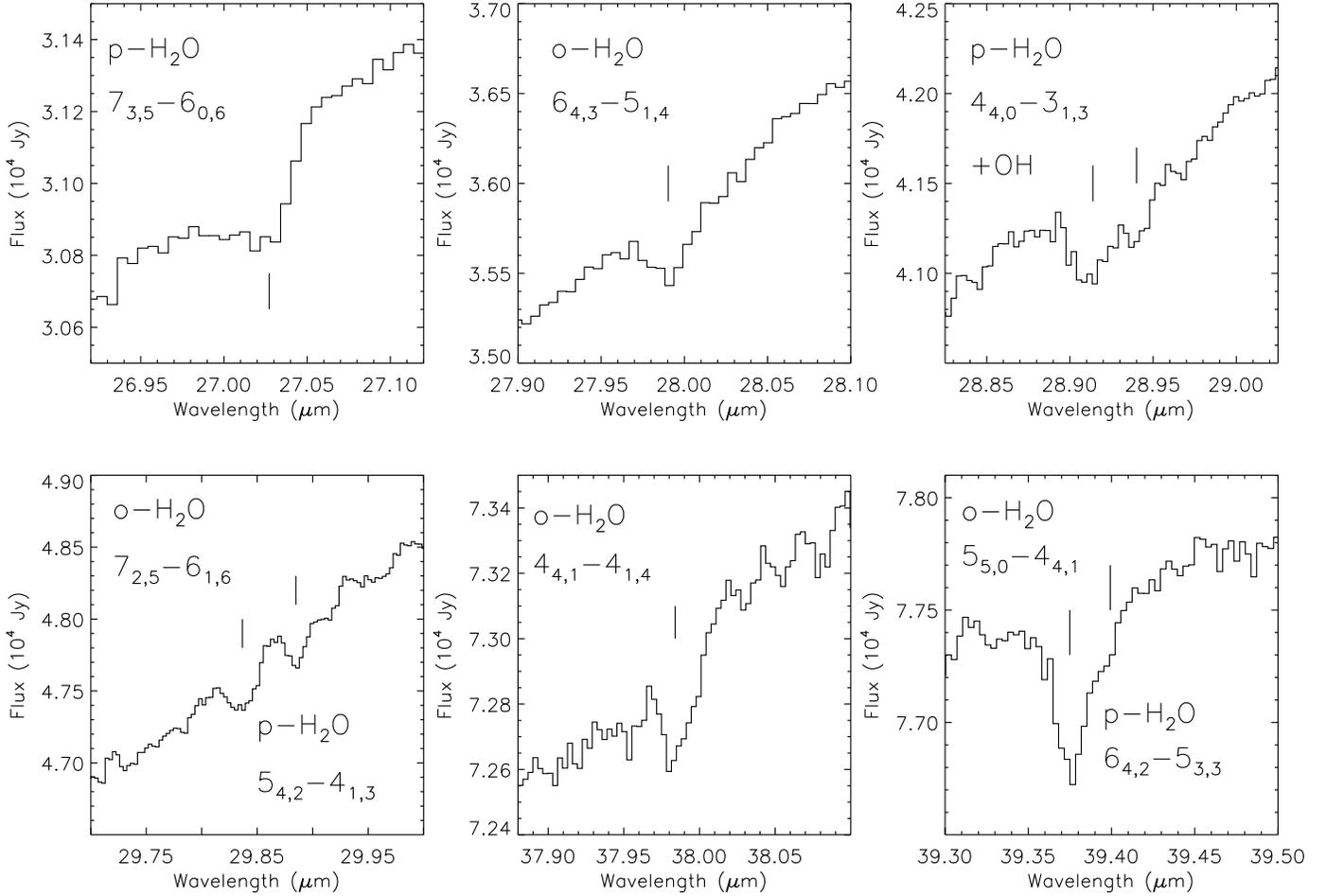}}
\vspace*{0.25cm}
\caption[]{Close-ups of several of the H$_{2}$O absorption lines shown
in Figure 1. The rest wavelength of each line is indicated by a tick mark.
The p-H$_2$O 6$_{42}$--5$_{33}$ 39.3992 $\mu$m line is visible
as a shoulder on the deeper o-H$_2$O 5$_{50}$--4$_{41}$ 39.3749 $\mu$m line.}
\end{figure*}

\begin{figure*}
\vspace*{-0.75cm}
\hspace*{2.0cm}
\resizebox{18cm}{!}{\includegraphics[angle=90]{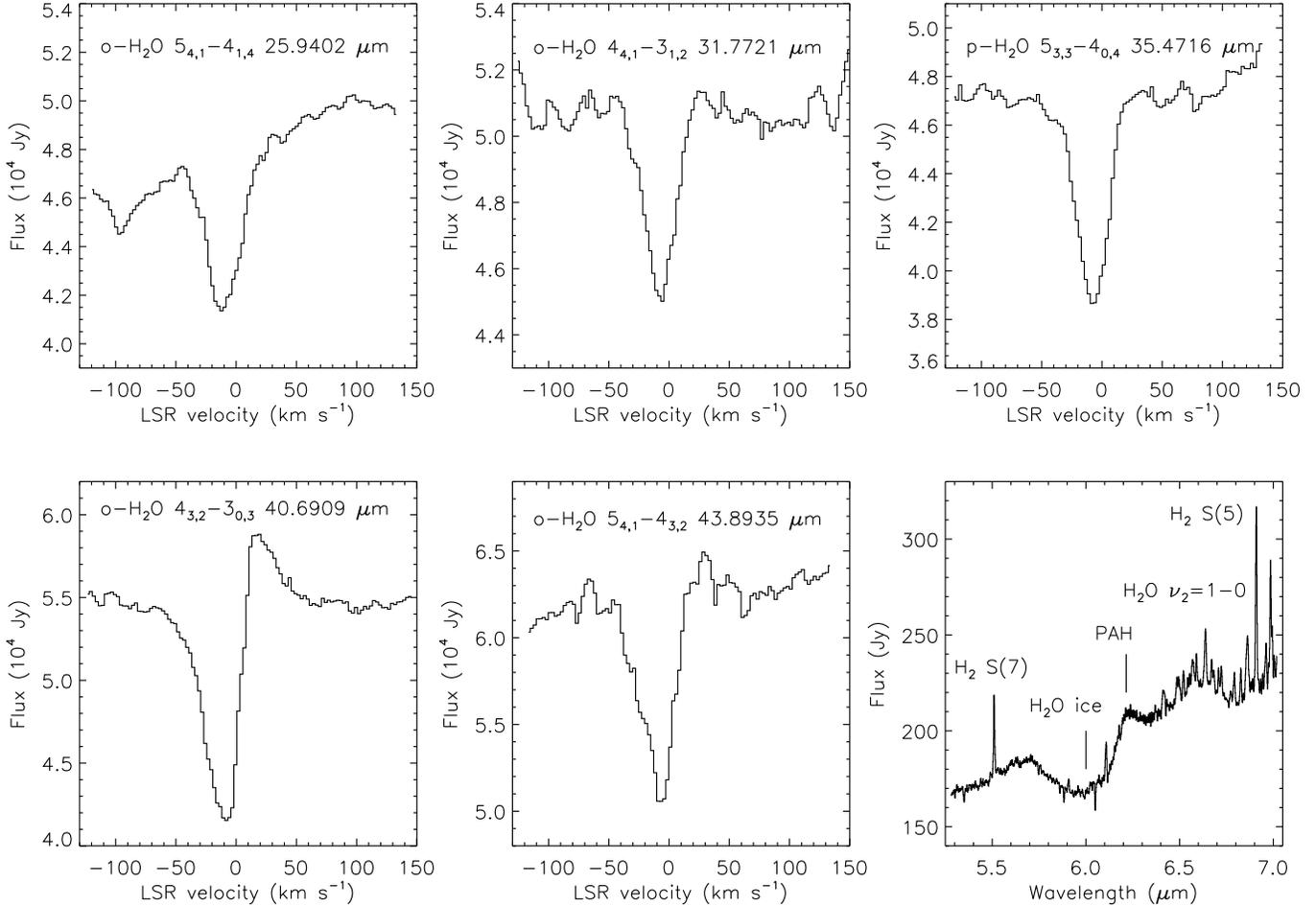}}
\vspace*{0.25cm}
\caption[]{ISO--SWS Fabry-Perot spectra toward IRc2 of a selection of 
H$_{2}$O absorption lines. The final panel displays the 5.3--7.0 $\mu$m 
grating spectrum of the H$_2$O $\nu_{2}$=1--0 band taken simultaneously 
with the revolution 831 Fabry-Perot spectra.}
\end{figure*}

\begin{figure}
\resizebox{\hsize}{!}{\includegraphics{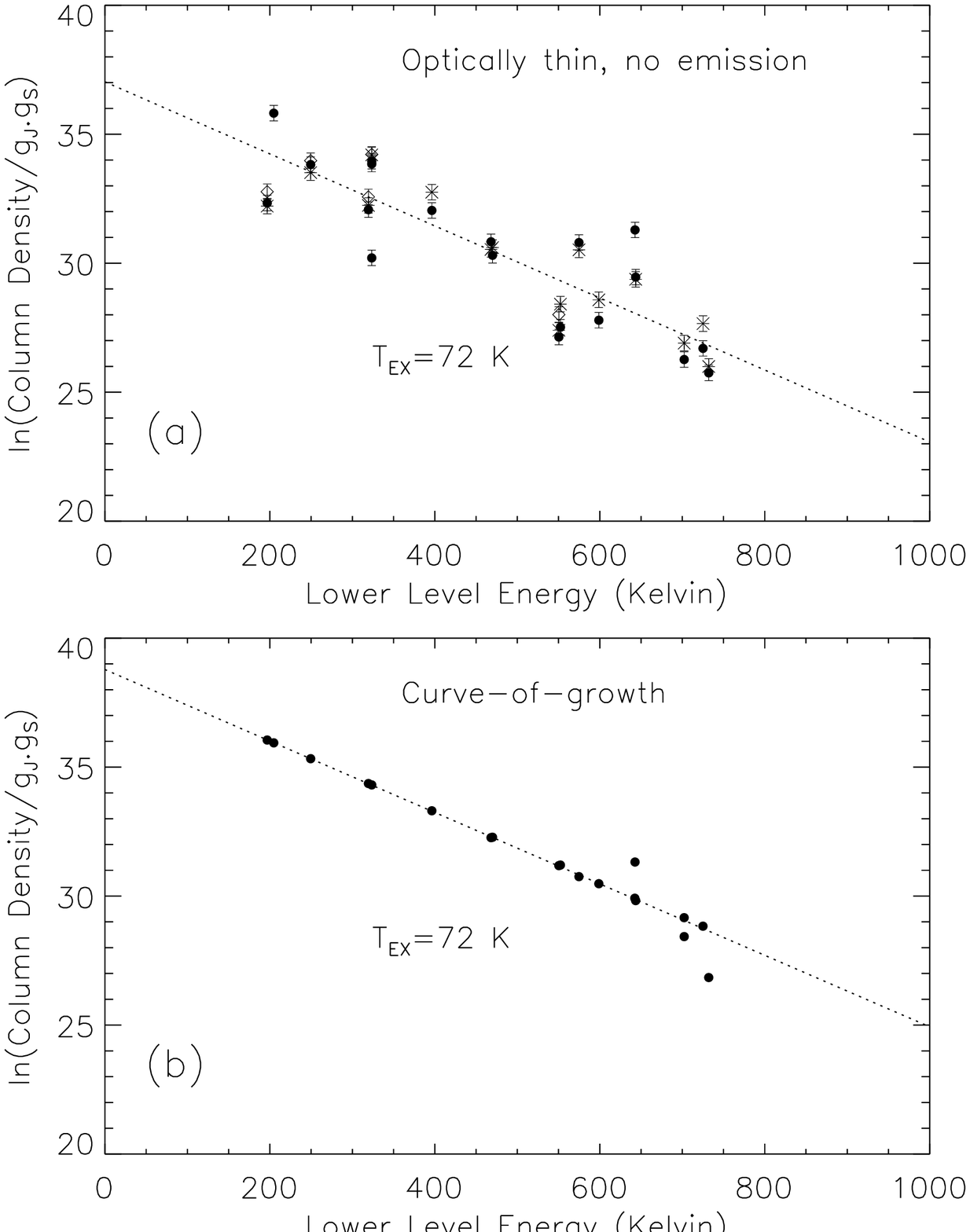}}
\caption[]{(a) Level populations of H$_2$O derived from the absorption lines
and using the optically thin formula 
$N_{\ell}=(8 \pi c W_{\lambda}g_{\ell})/(\lambda^{4}A_{u\ell}g_{u})$.
The solid circles represent the revolution 660 grating data, the asterisks 
the revolution 826 grating data, and the open diamonds the revolution 823 
and 831 Fabry-Perot data. The dotted line represents a best fit 1$^{\rm st}$ 
order polynomial to the revolution 660 data, and implies an excitation 
temperature of 71.6 K. A similar fit to the revolution 826 data yields 
75.9 K, whilst the Fabry-Perot absorption (emission) data yield 62.9 K 
(64.4 K).  
(b) Level populations of H$_{2}$O derived from the generalised
curve-of-growth (GCOG) analysis of the revolution 660 data. The best fit 
GCOG analysis for the populations results in a distribution that is 
consistent with a thermal distribution at a constant excitation 
temperature of 72.5~K. The dotted line indicates a best fit 1$^{\rm st}$ 
order polynomial to the points with $E_{\ell}$ $<$ 600 K.}
\end{figure}

Table 1 presents an overview of our observations for the Revolution
660 data, including the rest wavelength, upper and lower levels, and
equivalent widths $W=c\,W_{\lambda}/\lambda$ in units of Doppler velocity 
(km s$^{-1}$), where $W_{\lambda}$ is the line equivalent width in 
wavelength units. The superscripts on $W$ refer to the observed and
calculated equivalent widths, the latter derived from a generalised 
curve-of-growth analysis to be described in Sect. 4.2. Table~1 also 
contains the results for three pure rotational OH lines detected in 
absorption in our spectra. All derived quantities were calculated from 
unshifted spectral segments. Figure 4(a) presents our results graphically, 
in the form of a standard population diagram (Goldsmith \& Langer 1999),
where the column density in the lower level has been determined using the 
formula $N_{\ell}=(8 \pi c W_{\lambda}g_{\ell})/(\lambda^{4}A_{u\ell}g_{u})$, 
where $g_{\ell}$ and $g_u$ are the statistical weights of the lower and 
upper levels, $A_{u\ell}$ is the Einstein $A$-coefficient in s$^{-1}$ 
and $W_{\lambda}$ is the equivalent width. This formula is only strictly 
valid for optically thin lines, and under the assumption that the
upper levels are not significantly populated. The error bars represent
a 30\% uncertainty, so that the scatter in the data is clearly larger 
than any systematic or statistical uncertainties. Although there is a clear
trend in the data, this scatter already hints that a more detailed
analysis is required to obtain a reliable total H$_2$O column density.
Such an analysis, in the form of a generalised curve-of-growth, is
presented in Sect. 4.2.

Almost all lines listed in Table 1 at 27.5--45.2 $\mu$m are visible in both 
the Revolution 660 and 826 spectra, and their measured equivalent widths agree 
within 30\% in the majority of cases. See Fig. 4(a), where both data sets
are presented graphically. We present only the Revolution 660 data in Table 1
because the data from that revolution are of a better quality, and also only
15 of the 19 lines were observed in Revolution 826 due to the shorter 
wavelength scan range. The fact that the pure rotational H$_2$O and OH lines 
observed with the SWS occur in absorption is in strong contrast with the 
lines seen with the LWS, which are mostly in emission 
(e.g., Harwit et al.\ 1998). Since the LWS lines arise in levels with 
similar excitation energies at wavelengths where the continuum flux is 
comparable to that in the SWS range, this suggests that most of the 
mid-infrared continuum and H$_2$O absorption originate in a small region 
of size comparable to the SWS beam. In the almost order-of-magnitude 
larger area of the LWS beam, even more of the absorption is probably 
filled in by large-scale emission from more extended, lower excitation 
shocked gas. The qualitative change in the character of the water spectrum 
is thus probably a result of effects of both geometry (beam-filling) and 
radiative transfer.

The small differences in equivalent widths found from the grating and 
Fabry-Perot spectra are due to uncertainties in the calibration and 
varying beam shapes, which may couple differently to the complex source 
structure. The continuum radiation at 25--45 $\mu$m is produced by
extended dust emission over a $\sim$ 30$''$ region centered near IRc2, in 
contrast with the situation at $\sim 6$ $\mu$m where BN dominates
(Wynn-Williams et al.\ 1984). Since H$_2$O emission is observed on an 
arcmin scale with the LWS (e.g. Cernicharo et al. 1999a), it is likely 
that the H$_2$O covers all of 
the continuum, at least for the lower-excitation lines. However, the 
absorption of the lower-lying lines may be partially filled in by 
emission within the SWS beam. As discussed below, these effects must be 
taken into account explicitly in an analysis of the equivalent widths.

\begin{table*}
\caption[ ]{Revolution 660 ISO--SWS H$_{2}$O and OH observations toward
Orion--IRc2}
\begin{flushleft}
\begin{tabular}{llllllll}
\noalign{\smallskip}
\hline
\noalign{\smallskip}
Wavelength  &  Species  &  Transition  &  $hcE_{\ell}/k$  &  $W^{\rm obs}$  &
$W^{\rm GCOG}$  &  KN96$^{\rm a}$ & $\tau_{0}$ \\
(rest, $\mu$m)  &  &  $u - \ell$  &  (K)  &  (km s$^{-1}$)  & (km s$^{-1}$) & 
(km s$^{-1}$) & \\
\noalign{\smallskip}
\hline
\noalign{\smallskip}
\multicolumn{8}{c}{Grating data -- H$_{2}$O}\\
\hline
25.9402 & o-H$_{2}$O &  5$_{41}$--4$_{14}$ & 323.49 & 3.41 & 2.88 & 3.4 & 
0.16 \\
27.0272 & p-H$_{2}$O &  7$_{35}$--6$_{06}$ & 642.69 & 2.18 & 2.12 & 0.01 & 
0.08 \\
27.9903 & o-H$_{2}$O &  6$_{43}$--5$_{14}$ & 574.73 & 2.17 & 1.90 & & 0.07 \\
28.9138 & p-H$_{2}$O &  4$_{40}$--3$_{13}$ & 204.71 & 2.58 & 2.28 & 1.6 &
0.10 \\
29.8367 & o-H$_{2}$O &  7$_{25}$--6$_{16}$ & 643.49 & 1.44 & 1.35 & 0.5 &
0.07 \\
29.8849 & p-H$_{2}$O &  5$_{42}$--4$_{13}$ & 396.38 & 0.82 & 0.82 & & 0.10 \\
30.8994 & o-H$_{2}$O &  6$_{34}$--5$_{05}$ & 468.10 & 3.91 & 3.98 & 7.3 &
0.54 \\
31.7721 & o-H$_{2}$O &  4$_{41}$--3$_{12}$ & 249.43 & 2.81 & 2.83 & 5.3 & 
0.42 \\
35.4716 & p-H$_{2}$O &  5$_{33}$--4$_{04}$ & 319.48 & 3.53 & 3.61 & 7.7 &
1.13 \\
36.2125 & p-H$_{2}$O &  6$_{24}$--5$_{15}$ & 469.94 & 1.56 & 1.57 & 0.9 &
0.38 \\
37.9839 & o-H$_{2}$O &  4$_{41}$--4$_{14}$ & 323.49 & 0.79 & 0.11 & 1.0 & 
0.04 \\
39.3749 & o-H$_{2}$O &  5$_{50}$--4$_{41}$ & 702.27 & 1.45 & 1.60 & 0.05 & 
0.87 \\
39.3992 & p-H$_{2}$O &  6$_{42}$--5$_{33}$ & 725.09 & 0.52 & 0.52 &  & 0.14 \\
40.3368 & o-H$_{2}$O &  6$_{43}$--5$_{32}$ & 732.06 & 0.62 & 0.61 & 0.15 &
0.06 \\
40.6909 & o-H$_{2}$O &  4$_{32}$--3$_{03}$ & 196.77 & 7.26 & 7.18 & 29.8 &
9.67 \\
40.7597 & p-H$_{2}$O &  6$_{33}$--5$_{24}$ & 598.83 & 0.71 & 0.66 & 0.04 &
0.34 \\
43.8935 & o-H$_{2}$O &  5$_{41}$--4$_{32}$ & 550.35 & 2.56 & 2.63 & 3.7 &
4.74 \\
44.1946 & p-H$_{2}$O &  5$_{42}$--4$_{31}$ & 552.26 & 1.26 & 1.26 & 0.18 &
1.63 \\
45.1116 & o-H$_{2}$O &  5$_{23}$--4$_{14}$ & 323.49 & 6.42 & 6.77 & 30.9 &
12.8 \\
\hline
\multicolumn{8}{c}{Fabry-Perot} \\
\hline
25.9402 & o-H$_{2}$O & 5$_{41}$--4$_{14}$ & 323.49 & 4.44 & 2.88 & 3.4 & \\
31.7721 & o-H$_{2}$O & 4$_{41}$--3$_{12}$ & 249.43 & 3.24 & 2.83 & 5.3 & \\
35.4716 & p-H$_{2}$O & 5$_{33}$--4$_{04}$ & 319.48 & 5.72 & 3.61 & 7.7 & \\
40.6909$^{\rm b}$ & o-H$_{2}$O & 4$_{32}$--3$_{03}$ & 196.77 & 5.20
& 7.18 & 29.8 & \\
43.8935$^{\rm c}$ & o-H$_{2}$O & 5$_{41}$--4$_{32}$ & 550.35 &  4.48
& 2.63 & 3.7 & \\
\hline
\multicolumn{8}{c}{Grating data -- OH $^{2}\Pi_{1/2}$--$^{2}\Pi_{3/2}$, $N,J=$}
 \\
\hline
28.93461 & OH    & 4,$\frac{7}{2}e,f$--2,$\frac{5}{2}e,f$ & 120.46 & 2.86 &  &
 & \\
34.59617 & OH    & 3,$\frac{5}{2}f$--1,$\frac{3}{2}f$ & 0.08 & 4.46 &  &  & \\
34.62215 & OH    & 3,$\frac{5}{2}e$--1,$\frac{3}{2}e$ & 0.00 & 4.68 &  &  & \\
\noalign{\smallskip}
\hline

\end{tabular}
\end{flushleft}

\begin{list}{}{}
\item[$^{\rm a}$] Orion shock model of Kaufman \& Neufeld (1996), using 
lower oxygen and H$_2$O abundances of $3.16\times10^{-4}$ and 
$3.5\times10^{-4}$ respectively, cf.\ Harwit et al.\ (1998), and 
assuming a shock velocity of 37 km s$^{-1}$.
\item[$^{\rm b}$] The net equivalent width, including both emission and
absorption components. Individually, the observed absorption component is 
11.07 km s$^{-1}$, whilst the emission component flux is 
7.8$\times$10$^{-18}$ W cm$^{-2}$.
\item[$^{\rm c}$] The net equivalent width, including both emission and
absorption components. Individually, the observed absorption component is 
6.11 km s$^{-1}$, whilst the emission component flux is 
2.3$\times$10$^{-18}$ W cm$^{-2}$.
\end{list}

\end{table*}

\section{Analysis and discussion}
 
\subsection{Spatial origin of the water}
 
It is important to establish where the observed water is located
along the line-of-sight towards IRc2. Possible candidates are the
quiescent ridge, the shocked (low and/or high velocity) plateau gas,
the hot core, or a combination of these components. Fortunately, our 
Fabry-Perot observations can aid in resolving the ambiguity. For the 
3 pure absorption cases, the $v_{\rm LSR}$ is $\sim -8$$\pm3$
km s$^{-1}$. This blueshift is similar to velocities found for molecules
such as CO, C$_{2}$H$_{2}$ and OCS of $-3$ to $-18$ km s$^{-1}$ in absorption 
by Scoville et al.\ (1983) and Evans et al.\ (1991), which they interpret
as arising from the low-velocity plateau gas, representing a shell
expanding about a point close to IRc2. The most likely candidate for the
location of our H$_{2}$O gas is therefore the shocked, low-velocity plateau
gas. Our lines are also resolved with observed full-width at half-peak
$\Delta V \simeq$ 30 km s$^{-1}$. Assuming that the intrinsic source 
profile is Gaussian and that the FP instrumental profile is an ideal 
Airy profile with FWHM of 10 km s$^{-1}$, the intrinsic line widths are 
of order 23 km s$^{-1}$. This broad linewidth excludes most other 
components, such as the quiescent ridge and hot core, which have 
$\Delta V<20$ km s$^{-1}$ (Blake et al.\ 1987). H$_2$O differs in this 
respect from the other molecules seen in absorption, which have $\Delta V$ 
of order only 3--10 km s$^{-1}$ (Scoville et al.\ 1983, Evans et al.\ 1991). 

The peak of the emission component of the 40.69 $\mu$m line occurs at 
$v_{\rm LSR}$=+7 to $\sim$ +12 km s$^{-1}$. This velocity range is similar 
to that found by Cernicharo et al.\ (1994) for the $3_{1,3}-2_{2,0}$ 183 GHz 
line, as well as for a selection of H$_{2}^{18}$O lines by Zmuidzinas 
et al.\ (1995), and is in fact close to $v_{\rm LSR}\approx 9$~km s$^{-1}$ 
appropriate for the dense, quiescent ridge in which IRc2 is embedded. The 
emission component is also resolved, with an observed linewidth of 
$\Delta V \approx 40$ km s$^{-1}$, consistent with the intrinsic widths 
of $\sim$ 35 km s$^{-1}$ of the water lines detected in emission at longer 
wavelengths with the ISO--LWS by Harwit et al.\ (1998). 

In the comparisons presented above the absorbing and emitting components have
been considered separately. However, the profile as a whole for the 40.69
$\mu$m line, and to some extent also the 43.89 $\mu$m line, is very 
reminiscent of a P Cygni-type profile, and is similar to those observed for 
several far-infrared OH lines toward Orion by Betz \& Boreiko (1989) and 
Melnick et al.\ (1990). In fact, the total blue-shift of the absorption 
component for all 5 lines, i.e. with respect to the rest velocity of the 
cloud $v_{\rm LSR}\approx 9$~km s$^{-1}$, is very close to the expansion 
velocity of the ``low-velocity flow'', (or plateau, or ``expanding
doughnut'') of 18$\pm$2 km s$^{-1}$ quoted by Genzel et al.\ (1981) from
proper motion studies of the 22 GHz H$_{2}$O maser emission. Together with the
fact that the emission component of the 40.69 $\mu$m line is near the 
rest velocity of the cloud, this is precisely the definition of a P-Cygni 
profile and establishes almost without a doubt that the observed water 
arises from an outflow centered near IRc2, most likely from source ``I'' 
(Menten \& Reid 1995). As noted by Genzel et al.\ (1981) this outflow, or 
low velocity plateau, can be traced from $\sim2\times10^{17}$ cm (30$''$) 
to within 10$^{14}$ cm from its dynamical center. 

We note that Takahashi et al. (1985) predicted that many 
H$_2$O lines from a linearly expanding spherical cloud would exhibit P-Cygni 
type profiles, although the only case in common with our observations, the 
40.69 $\mu$m line, is not P-Cygni shaped in any of their models. However, 
they do not include shock emission in their calculations. Further, there is
a difference between a true P-Cygni profile and the P-Cygni {\em type} 
profiles observed here and predicted by Takahashi et al. A true P-Cygni 
profile results from an expanding stellar atmosphere {\em with a central 
luminosity source}, whilst the P-Cygni type profiles of far-infrared water 
lines are caused by the presence of warm dust {\em throughout} the expanding 
cloud. Further details on water line profiles in regions of intense dust
emission may be found in Takahashi et al. (1985) or Doty \& Neufeld (1997).
 
\subsection{Excitation of the absorbing water vapour}

\subsubsection{Generalised curve-of-growth method}

The ISO spectra of Orion highlight the need for a non-traditional analysis
of interstellar spectra. At shorter wavelengths (in the visible and 
ultraviolet), molecular absorption lines arise in gas that is physically
very cold in relation to the radiation temperature of the background star,
so that the absorption can be described with a classical curve-of-growth
analysis and the effects of stimulated emission can be neglected. At 
longer wavelengths (radio), an isolated molecular region typically exhibits
collisionally excited emission lines on top of a weak continuum so 
that the most significant radiative coupling is to the cosmic background
radiation at $T_{\rm cbr}=2.728$ K. In the present case, we observe a 
molecular spectrum that is formed in the presence of a very strong continuum,
i.e., where the radiation temperature of the continuum may be comparable to
the excitation energies of the states involved. This implies both that the
classical curve-of-growth analysis of cold absorbers detached from the 
background continuum is invalid and that the standard analysis of 
collisional excitation in competition with a very cold, dilute background
is inadequate. Indeed, it is the near equality of continuum
brightness temperature $T_{\rm rad}(\nu)$ and molecular excitation temperature
$T_{\rm ex}$ that explains at least partly why the water spectrum of 
Orion IRc2 is predominantly in absorption at $\lambda \la 50\;\mu$m 
and in emission at longer wavelengths. In detail,
such spectra should be modelled as an extended atmosphere in which lines
and continuum are treated consistently. We will outline briefly how it
is possible to extract important information on excitation and abundance
through a simpler analysis of the unresolved lines based on a generalised
curve-of-growth.  Details of this method will be 
discussed elsewhere (Black, in preparation).

Consider the situation in which a fractional area $a_c$ of a
continuum source of intensity $I_c(\nu)$ is obscured by a column of gas
in the foreground.  The observed spectrum in the vicinity of a transition
$u \leftrightarrow \ell$ is given by the sum of the unattenuated continuum
(first term), the absorbed continuum (second term), and the emission of
the cloud itself (third term):
\begin{eqnarray}
I_{\rm obs}(\nu) & = & (1-a_c)I_c(\nu) \nonumber \\
              & + & a_c I_c(\nu) e^{-\tau_{u\ell}(\nu)} \nonumber \\
  & + & B_{\nu}(T_{u\ell}) \Bigl(1 - e^{-\tau_{u\ell}(\nu)} \Bigr) \;\;\; ,
\end{eqnarray}
where $\tau_{u\ell}$ is the optical depth and
$B_{\nu}(T_{u\ell})$ is the Planck function evaluated for an
excitation temperature $T_{u\ell}$. This temperature is defined by the
column densities of molecules in the upper ($N_u$) and lower 
$(N_{\ell})$ states through:
$$ {{N_u}\over{N_{\ell}}} = {{g_u}\over{g_{\ell}}} \exp\Bigl(-{{h\nu}\over
{kT_{u\ell}}}\Bigr) \;\;\;. \eqno(2) $$
The optical depth function for a single line can be written (e.g. Rybicki \& 
Lightman 1979, Wannier et al. 1991)
$$ \tau_{u\ell}(\nu) = 3.738\times 10^{-7} 
{{A_{u\ell}}\over{\tilde{\nu}_{u\ell}^3}}
{{N_{\ell}}\over{\Delta V}} {{g_u}\over{g_{\ell}}} \Bigl(1 -
\exp\bigl(-{{h\nu}\over{kT_{u\ell}}}\bigr) \Bigr) \phi(\nu) \eqno(3) $$
where 
$N_{\ell}$ is in cm$^{-2}$, $\tilde{\nu}_{u\ell} = \nu_{u\ell}/c$ is the
wavenumber of the line in cm$^{-1}$, and the line shape function is
a gaussian of full-width at half-maximum $\Delta V$ in km s$^{-1}$,
normalized so that $\phi(\nu_{u\ell}) = 1.0$\ .
The term in parentheses is the correction for stimulated emission. 
 
The generalised equivalent width, in frequency units, is expressed as
$$ W_{\nu}^{\rm obs} = \int {{I_c(\nu) - I_{\rm obs}(\nu)}\over{I_c(\nu)}}
d\nu \eqno(4a) $$
$$ W_{\nu}^{\rm obs} = \int \Bigl[\Bigl( a_c -{{B_{\nu}(T_{u\ell})}\over
{I_c(\nu)}}\Bigr)\Bigl(1 - e^{-\tau_{u\ell}(\nu)}\Bigr)\Bigr] d\nu \;\;\;.
\eqno(4b) $$
In the limit as $a_c\sim 1$ and $T_{u\ell}<< h\nu/k$,
the equivalent width approaches the classical form of cold absorbers 
detached from the background continuum. By convention,
the equivalent width is positive for net absorption and negative
for emission. Emission can arise either when $T_{u\ell}$ is large enough
that $B_{\nu}(T_{u\ell})/I_c(\nu) > a_c$ or when there is a population
inversion ($N_u/N_{\ell} > g_u/g_{\ell}$) so that $\tau_{u\ell}$ is negative,
as in a maser.
 
For a single unresolved line of equivalent width $ W_{\nu}^{\rm obs}$, there
is not a unique solution, because 4 parameters must be constrained:
$a_c$, $N_u$, $N_{\ell}$, and $\Delta V$. In the classic problem, in
which a continuum point source of intensity is located behind, and completely
obscured by, a column of cold foreground gas, and there is negligible 
population in the upper level of an absorption transition, then $a_c=1$ 
and $N_u/N_{\ell}\sim 0$ respectively. In that case, two lines arising 
in a common level suffice to determine $N_{\ell}$ and $\Delta V$. 

\subsubsection{Results of the generalised curve-of-growth analysis}

We have analyzed the equivalent widths using Eq. (4b). If we
take the continuum intensity to be approximately constant over the line
profile, then the term $a_c -B_{\nu}(T_{u\ell})/I_c(\nu)$ can be taken outside
the integral in Eq. (4b). We adopt an empirical continuum intensity 
$I_c(\nu) = f_{\nu}^{\rm obs}/\Omega_{\rm beam}$, where $f_{\nu}^{\rm obs}$
is the flux density observed in the ISO--SWS aperture of solid angle 
$\Omega_{\rm beam}$. When $I_c(\nu)$ is fixed in this way, the equivalent 
width of a line depends on the parameters $N_{\ell}$, $T_{u\ell}$, 
$\Delta V$, and $a_c$. We adopt an intrinsic width $\Delta V=23$ km s$^{-1}$
and assume that $a_c$ has the same value for
all lines in the following analysis. Thus each line is characterized
by two parameters $N_{\ell}$ and $T_{u\ell}$ (or $N_{\ell}$ and $N_u$).
The solution is still underdetermined because there are 29 distinct energy 
states involved in the 19 observed transitions. Fortunately, there are 
several cases in which some of the same states are involved in two or 
more transitions. As a first approximation, it is assumed that a single 
value of the excitation temperature $T_{u\ell}=T_{\rm ex,0}$ applies to
all transitions with lower-state energies $E_{\ell}\leq 275$ cm$^{-1}$.
The computed equivalent widths (Eq. 4b) agree with the observed values 
within 33\%\ for all but the weakest one of these lines, when 
$a_c=1$, $T_{\rm ex,0}=72.5$ K, and $N_{\rm total}=1.5\times 10^{18}$ 
cm$^{-2}$. This solution is not very sensitive to the value of the continuum
covering factor: for $a_c=0.9$, we obtain $T_{\rm ex,0}=71.0$ K and 
$N_{\rm total}=1.6\times 10^{18}$ cm$^{-2}$, while for $a_c=0.8$, the 
corresponding results are 69.1 K and $1.7\times 10^{18}$ cm$^{-2}$.

With the populations of the lowest energy states ($E_{\ell}\leq 275$ 
cm$^{-1}$) fixed at $T_{\rm ex,0}=72.5$ 
K, the column densities in higher states are adjusted until the standard
deviation of the mean $W^{\rm obs}-W^{\rm GCOG}$ is minimized for the 
entire set of lines. In this solution, the observed $W^{\rm obs}$ and 
calculated $W^{\rm GCOG}$ equivalent widths all agree within 16\%, with the 
single exception of the $4_{41}-4_{14}$ line at 37.98 $\mu$m. 
The calculated equivalent widths are listed in Table 1. Note that the
adjustment also improves the fit for the lines arising in the low-lying
states because $W^{\rm GCOG}$ is sensitive to column densities in both the
upper and lower states. The result is a population distribution that is
indistinguishable from a thermal distribution at 72.5 K for all states with
$hcE_i/k\lse 700$ K. A few of the observed levels at higher energies 
($hcE_i/k > 700$ K) are well constrained because they are the 
upper states of more than one transition, and their populations 
start to clearly show effects of subthermal excitation, 
$T_{u\ell} \approx 55$ K. Even though the adopted linewidth is large, 
the derived line-center optical depths, listed in Table 1, reach 
values as high as $\tau_0=12.8$ and 9.7 in the 45.11 and 40.69 $\mu$m 
lines, respectively. The smallest values of optical depth are 
$\tau_0\approx 0.06$. Finally, ortho and para states of H$_2$O are 
consistent with the same excitation temperature and with an ortho/para 
ratio of 3.

The results of our generalised curve-of-growth analysis are shown
graphically in Fig. 4(b), in the form of a standard population
diagram.  The much reduced scatter in the data is clearly a marked
improvement over the optically thin approach assuming no stimulated
emission used in obtaining Fig. 4(a). It is however interesting that
the inferred excitation temperatures from the two approaches are
similar. The excitation temperature of $\sim$ 72 K is similar to the
colour temperature of the dust between 20 and 100 $\mu$m inferred by
Werner et al.\ (1976), which confirms the expectation that pumping by
infrared continuum radiation from dust plays an important role in the
H$_{2}$O excitation. In this case, there is no direct constraint on
the kinetic temperature of the H$_2$O-containing gas. Subthermal
excitation by collisions at densities below the critical densities of
the observed levels ($\sim 10^9$ cm$^{-3}$) can also result in
excitation temperatures of order 50--100~K, but this merely
re-affirms that radiative processes probably dominate the excitation.
We also note that the excitation temperature inferred from the
emission components of the 40.69 and 43.89 $\mu$m lines is $\sim$ 64
K, also implying a large contribution by radiative processes to the
populations.  Moreover, the analysis suggests that $T_{u\ell}$ of the
absorption lines is so close to the continuum radiation temperature in
this wavelength range, that it is easy to see why the H$_2$O spectrum
should go into emission at $\lambda > 50\;\;\mu$m as the opacity and
radiation temperature of the continuum decrease with increasing
wavelength.

Because the lowest state involved in the 25--45 $\mu$m spectrum lies at
$hcE_i/k=197$ K, we cannot determine the column densities in the most
populated states directly from these observations. However, if we assume
that the observed excitation pattern applies to the unobserved lower states, 
then we infer a total column density of 
$N$(H$_2$O) = $1.5\times 10^{18}$ cm$^{-2}$ averaged over the ISO--SWS 
beam toward Orion IRc2. 

The net absorption by water in Orion is sensitive to the competition between
the molecular excitation and the continuum brightness. We have assumed that 
the true continuum intensity in Eq. (4) is equal to the mean surface 
brightness observed in the spectrum (flux averaged over the aperture). In 
general this need not be true, and the analysis can include an additional
correction factor for the coupling to the continuum. Tests show that if the
continuum intensity is thus changed by a factor of two in either direction,
the derived total column density and mean excitation temperature change by 
less than 20\%, which we take as our maximum uncertainty. 

\subsubsection{Shock models}

We have also compared our observed line equivalent widths with those computed
for the Kaufman \& Neufeld (1996) C-shock in Orion with a pre-shock density
of 10$^{5}$ cm$^{-3}$ and shock velocity of 37 km s$^{-1}$ (Kaufman \&
Neufeld, private communication) (see Table 1). The water abundance (with
respect to H$_{2}$) used in this shock calculation was 3.5$\times$10$^{-4}$,
based on the O and C abundances of Cardelli et al.\ (1996), with all the
carbon locked up in CO and H$_{2}$O accounting for all the remaining oxygen.
Although this model refers to ``peak 1'' rather than the IRc2 region, it
reproduces the equivalent widths quite well within a factor of a few, except
for some of the higher-lying lines (e.g., 27.03, 39.37, 40.76, 44.19 $\mu$m).
Most of these lines are likely to be strongly affected by radiative
excitation, which was not included in their model. 

We note that the shock velocity used is 37 km s$^{-1}$, different from
the expansion velocity inferred from our FP observations of 18 km s$^{-1}$. 
Indeed, our modelling of our H$_2$ emission, using the Kaufman \& Neufeld 
(1996) shock model, indicates that at least 2 shock components are required. 
One component at 15--20 km s$^{-1}$ matches well the observed column 
densities in the $v$=0, $J$=3 to 7 levels, whilst a second component, at 
35 km s$^{-1}$, is required to match higher $J$ levels and the ro-vibrational 
lines (Wright 1999). Harwit et al. (1998) used a shock velocity of 
37 km s$^{-1}$ to match their 8 LWS water line emission detections from 72 
to 125 $\mu$m. These data were taken at a slightly different position,
centred on BN rather than IRc2 and offset 8.2$''$ to the north. The SWS 
results may therefore be more sensitive to the low-velocity plateau gas 
identified by several authors (e.g. Genzel \& Stutzki 1989), whereas the 
larger LWS beam will include a significant contribution from the high-velocity
gas. We also note that water line emission in the Kaufman \& Neufeld (1996) 
shock models is for many lines insensitive to the shock velocity, especially 
when the velocity is $\geq$ 15--20 km s$^{-1}$. This is presumably because 
in such cases the water abundance has reached its maximum value and that 
the gas temperature is well above the upper levels involved. Therefore, the 
emission depends more on density and not temperature (which in turn is 
highly sensitive to the shock velocity). Even so, future attempts to match 
both the Harwit et al. lines and our detections with a slower shock, but 
including radiative excitation, will be worthwhile.

A long standing problem in shock research is the contribution of water
to the total gas cooling behind the shock, and it is pertinent to mention
it here. Unfortunately, the situation in Orion IRc2, i.e. a 
high water abundance (see Sect. 4.3) coupled with an intense mid- and
far-infrared radiation field, has not previously been considered in shock 
models. However, the detection of all our transitions in absorption raises 
the interesting possibility that at least these transitions are not cooling 
the gas, but rather heating it. Such heating results from the water molecules 
absorbing the radiation and thereby being raised to excited states. The 
excess energy may then be imparted to the gas through collisional 
de-excitation with other molecules, e.g. H$_2$, as in the scenario proposed 
by Takahashi et al. (1983). This would of course be in 
competition with radiative de-excitation (i.e. cooling), and is therefore 
density dependent. Further, Takahashi et al. (1983) and Neufeld et al. 
(1995) state that a necessary condition for H$_2$O heating is that the dust 
temperature is greater than the gas temperature. A full scale shock model, 
taking into account all transitions, would be required to determine if there 
was net heating or cooling. We merely mention it here as an interesting aside,
and direct the reader to the paper by Harwit et al. (1998) for a full 
discussion of the water cooling contribution in Orion. Briefly though, from
their larger beam LWS observations they find that H$_2$O and
CO contribute similar amounts to the overall cooling, but both are only 
about a tenth of the total H$_2$ cooling. On the other hand, Saraceno et al.
(1999), in a study of a sample of shock sources, find that CO cooling
typically dominates over H$_2$O by a factor of several.

\subsection{H$_2$O abundance in Orion}
 
In order to determine an H$_2$O abundance directly from our data, information
on the H$_2$ column density is needed. As noted above, from the H$_2$ 
lines detected elsewhere in our spectra (van Dishoeck et al.\ 1998), the 
shock has been characterized using the Kaufman \& Neufeld (1996) C-shock 
models, in a manner similar to that applied by Wright et al.\ (1996) to 
Cepheus A. Further details can be found in Wright (1999); in short, the 
H$_2$ pure rotational lines originating from $J$=3 to $J$=7 indicate a column 
$N$(H$_2$)=$(1.2\pm 0.2)\times 10^{21}$ cm$^{-2}$ of warm ($T_{ex}$=700 K) 
gas, and the inferred shock velocity of 15--20 km s$^{-1}$ is similar 
to the 18 km s$^{-1}$ expansion velocity deduced from our FP observations, 
but lower than that obtained from modeling the H$_2$ vibration-rotation lines 
(35 km s$^{-1}$). However, supporting evidence for the 
co-location of the H$_{2}$O and shocked H$_{2}$ comes from Scoville et al.\ 
(1982), who find that at and around the position of IRc2 the peak emission 
of the 1--0 S(1) line is at $v_{\rm LSR}$=$-6$ to $-16$ km s $^{-1}$, 
similar to the velocity of maximum H$_2$O absorption in our spectra. 

Using the above H$_2$ column density, and that of H$_2$O calculated
from the generalised curve-of-growth analysis, the inferred water abundance 
would be $1.25\times10^{-3}$. However, the inferred column density from 
the high-$J$ (i.e. 3--7) lines is a lower limit on the total column density, 
since the bulk of the H$_2$ is likely to be cooler than $T_{ex}$=700~K. 
Such cooler gas would be probed by the 28.2 $\mu$m H$_2$ 0--0 S(0) line,
but which unfortunately has not been detected toward IRc2, due to the 
extremely high continuum emission. However, a lower limit implies that 
$T_{ex}\geq 110$~K, calculated from the populations in $J$=3 and 2, 
inferred from the observed 0--0 S(1) and upper limit S(0) lines respectively.
On the other hand, the S(0) line is detected at about the 3$\sigma$ level 
toward the shock ``peak 2'', $\sim$ 30$''$ to the south-east of IRc2, and 
in this case the excitation temperature between the $J$=3 and $J$=2 levels 
is $\sim$ 150 K (Wright 1999). 

This temperature of $\sim$150 K is likely to be indicative of the 
kinetic temperature, $T_{kin}$, at this position, since the density of 
$\geq$10$^{5}$ cm$^{-3}$ (e.g. Genzel \& Stutzki 1989) is
above the critical density of the H$_2$ 0--0 S(0) and S(1) transitions
(e.g. Le Bourlot et al. 1999). Although this estimate of $T_{kin}$ is at 
a different position than where the water absorption line detections were 
made, Wright (1999) shows that in at least 3 shock sources, including Orion, 
the excitation temperature determined from pure rotational H$_2$ lines 
($J_{up}$=3--7) is quite invariant across the sources, as is the column 
density. Other estimates of $T_{kin}$ in the low-velocity plateau, 
tabulated by Genzel \& Stutzki (1989), range from 100--500 K.  

If we assume that $T_{kin}$ is between 100--500 K, then Table 2 
presents the expected total H$_2$ column density based on the observed
surface brightness of the 0--0 S(1) line toward IRc2 of 1.5$\times10^{-10}$ 
W cm$^{-2}$ sr$^{-1}$. We believe that 100 K is too low to be applicable to 
the shocked plateau gas, since it is lower than $T_{kin}$ we find at 
``peak 2'', and in any case would imply a column density an order of 
magnitude larger than the value of 10$^{23}$ cm$^{-2}$ quoted by Genzel \& 
Stutzki (1989). Also shown in Table 2 is the inferred water abundance, 
$N({\rm H_{2}O})/N({\rm H_{2}})$, using the observed water column density of 
1.5$\times$10$^{18}$ cm$^{-2}$, and assuming the water and molecular 
hydrogen to be co-located. Taking the maximum possible water abundance to 
be $5\times10^{-4}$ (anything higher would violate the constraint
imposed by the oxygen abundance itself of [O]/[H]=3.19$\times 10^{-4}$ 
from Meyer et al.\ 1998), then the water abundance is in the range of 
2.6--50$\times10^{-5}$.

\begin{table}
\caption[ ]{Inferred total H$_2$ column densities and water abundances}
\begin{flushleft}
\begin{tabular}{lll}
\noalign{\smallskip}
\hline
\noalign{\smallskip}
$T_{kin}$  &  $N({\rm H_{2}})^{\rm a}$  &  
$N({\rm H_{2}O})/N({\rm H_{2}})^{\rm b}$ \\
(Kelvin)  &  (cm$^{-2}$)  &  \\
\noalign{\smallskip}
\hline
\noalign{\smallskip}
100 & 1.1$\times$10$^{24}$ &  1.4$\times$10$^{-6}$  \\
150 & 5.7$\times$10$^{22}$ &  2.6$\times$10$^{-5}$  \\
200 & 1.4$\times$10$^{22}$ &  1.1$\times$10$^{-4}$  \\
230 & 8.1$\times$10$^{21}$ &  1.9$\times$10$^{-4}$  \\
250 & 6.2$\times$10$^{21}$ &  2.4$\times$10$^{-4}$  \\
280 & 4.4$\times$10$^{21}$ &  3.4$\times$10$^{-4}$  \\
300 & 3.7$\times$10$^{21}$ &  4.1$\times$10$^{-4}$  \\
320 & 3.2$\times$10$^{21}$ &  4.7$\times$10$^{-4}$  \\
350 & 2.6$\times$10$^{21}$ &  5.8$\times$10$^{-4}$  \\
400 & 2.1$\times$10$^{21}$ &  7.1$\times$10$^{-4}$  \\
450 & 1.7$\times$10$^{21}$ &  8.8$\times$10$^{-4}$  \\
500 & 1.5$\times$10$^{21}$ &  1.0$\times$10$^{-3}$  \\
\noalign{\smallskip}
\hline

\end{tabular}
\end{flushleft}

\begin{list}{}{}
\item[$^{\rm a}$] Assuming that the observed H$_2$ 0--0 S(1) 17.0348 $\mu$m 
line, with surface brightness 1.5$\times$10$^{-10}$ W cm$^{-2}$ sr $^{-1}$, 
arises from gas with kinetic temperature T$_{kin}$.
\item[$^{\rm b}$] Using a total water column density of 1.5$\times$10$^{18}$
cm$^{-2}$, and assuming the H$_2$ and H$_2$O are co-located.
\end{list}

\end{table}

Let us instead assume {\em a priori} that our water detections arise only
in that portion of the shock where the temperature is high enough for the
water creation reactions to rapidly proceed. Several authors (e.g. Hollenbach
\& McKee 1979, Neufeld et al. 1995, Ceccarelli et al. 1997 and Charnley 1997) 
have shown that water begins to be quickly formed when
the temperature exceeds about 230 K, and has reached its maximum abundance 
by about 320 K. In that case our inferred water abundance is in the range 
2--5$\times10^{-4}$, in remarkable agreement with the result of Harwit et al. 
(1998) of 3.5--5$\times$10$^{-4}$ considering the different methods used. 
Further, it is interesting to note that Watson et al. (1985) find a H$_2$ 
column density of $3\times10^{21}$ cm$^{-2}$, derived from several observed 
far-infrared CO lines in a 44$''$ beam, implying a H$_2$O abundance of 
5$\times$10$^{-4}$. Watson et al. however infer a kinetic temperature 
of 750 K, close to the excitation temperature derived from our H$_2$ 
data (Wright 1999), whereas from Table 2 their column density would instead 
imply a temperature of only $\sim$330 K.

Alternatively, if we simply use a H$_2$ column density of 10$^{23}$ cm$^{-2}$ 
for the plateau gas, as given by Genzel \& Stutzki (1989), then the water 
abundance is $1.5\times10^{-5}$, in better agreement with that of other 
shocks (e.g. Liseau et al.\ 1996, Ceccarelli et al.\ 1998, Spinoglio 
et al.\ 1999), and suggesting that a substantial fraction of the 
gas-phase oxygen may still be in another form, most likely gas-phase atomic 
O\,{\sc i}. However, this is probably a strict lower limit, since 
such a large column density probably includes gas at a temperature lower 
than 230 K, as well as gas behind the infrared continuum source.

An H$_2$O abundance at the high end of the range quoted above, i.e.
2--5$\times$10$^{-4}$, is favored if we assume that water is only 
efficiently created in regions where the temperature is greater than 
about 230 K. This water abundance is consistent with all of the gas-phase 
oxygen not locked up in CO being incorporated into H$_2$O. However, H$_2$O 
ice evaporates at temperatures above about 90 K, so lower temperatures, 
and hence a lower abundance, cannot be excluded. The major uncertainty 
on our abundance value is the total hydrogen column density.
As mentioned above, our result is in good agreement with Harwit et al.
(1998), but also in agreement with Cernicharo et al. (1999a,b) from 
LWS and ground based observations respectively. From radiative transfer 
modelling of their water detections, they find toward the Orion plateau a 
water abundance of order 1--2$\times$10$^{-4}$, at most a factor of a few
lower than what we find. Besides Orion, there is only one other source so 
far observed by ISO with such a high water abundance, namely L1448-mm by 
Nisini et al. (1999), with an abundance of order $5\times10^{-4}$.

A similar treatment for the OH lines detected in our spectrum yields a 
column density of approximately $5\times10^{16}$ cm$^{-2}$, after fixing 
T$_{ex}$ to be the same as that for water. The implied OH abundance is 
then $(8.8-156)\times10^{-7}$ for T$_{kin}$=150--320 K, or 
$(6.2-15.6)\times10^{-6}$ for 230--320 K.

\section{Conclusions}
 
Through use of the ISO--SWS in its grating mode, 19 pure rotational 
absorption lines of water have been detected toward Orion IRc2 for the 
first time. Fabry-Perot spectra of 5 lines reveal that the water is 
located in an outflow expanding at a velocity of 18 km s$^{-1}$. The 
strong mid-infrared continuum toward IRc2 plays a dominant role in the 
excitation of the molecule and the line formation, which can be modeled 
using a simple, generalised curve-of-growth technique. This yields a 
total water column density of order 1.5$\times10^{18}$ cm$^{-2}$ and 
excitation temperature of 72 K, similar to the dust continuum colour 
temperature. Both derived quantities have a maximum uncertainty of about
20\%. The data provide support for large abundances of H$_2$O 
in the outflows of massive stars. Simultaneous analysis of the complete 
ISO--SWS and LWS data set on H$_2$O, OH and CO may provide further 
information on the abundance and excitation of these molecules in the 
various physical components within the complex Orion environment.
 
\begin{acknowledgements} 
The authors are grateful to the SWS instrument teams in Groningen and
Garching and to the SIDT in Vilspa for making these observations possible.
They are indebted to G.A. Blake, A. Boonman, F. van der Tak, G.J. Melnick, 
A.G.G.M.\ Tielens and R. Timmermann for useful discussions. They are 
especially grateful to D.A. Neufeld and M.J. Kaufman for providing them 
with the equivalent widths of the H$_2$O lines in their models, and to 
W.F. Thi for developing the code for the up-down scan correction. This 
work was supported by NWO grant 614.41.003 and by the Spanish DGES grant 
PB96-0883 and PNIE grant ESP97-1618-E. During the final phases of this 
work CMW was supported by an ARC Australian Postdoctoral Research Fellowship.
\end{acknowledgements}


\begin{thebibliography}{}
 
\bibitem[]{}
Beckwith S., Persson S.E., Neugebauer G., Becklin, E.E., 1978,
ApJ 223, 464
 
\bibitem[]{}
Betz A.L., Boreiko R.T., 1989, ApJ 346, L101
 
\bibitem[]{}
Blake G.A.\ 1997, in: Molecules in Astrophysics: Probes and Processes,
IAU Symposium 178, ed.\ E.F.\ van Dishoeck (Dordrecht: Kluwer), p.31.
 
\bibitem[]{}
Blake G.A., Sutton E.C., Masson C.R., Phillips T.G., 1987,
ApJ 315, 621
 
\bibitem[]{}
Cardelli J.A., Meyer D.M., Jura M., Savage B.D., 1996, ApJ 467, 334
 
\bibitem[]{}
Ceccarelli C., Hollenbach D.J., Tielens A.G.G.M., 1997, ApJ 471, 400

\bibitem[]{}
Ceccarelli C., Caux E., White G.J., et al., 1998, A\&A 331, 372
 
\bibitem[]{}
Cernicharo J., Gonz\'alez-Alfonso E., Alcolea J., Bachiller R.,
John D., 1994, ApJ 432, L59
 
\bibitem[]{}
Cernicharo J., Lim T., Cox P., et al., 1997a, A\&A 323, L25
 
\bibitem[]{}
Cernicharo J., Gonz\'alez-Alfonso E., Lefloch B., 1997b, in:
First ISO workshop in Analytical Spectroscopy, eds.\ A.\ Heras
et al.\, ESA SP--419 (Noordwijk: ESTEC), p.\ 23
 
\bibitem[]{}
Cernicharo J., Gonz\'alez-Alfonso E., Sempere M.J. et al., 1999a, in: 
The Universe as seen by ISO, eds. P.Cox \& M.F. Kessler, ESA 
Publications Division SP-427, (Noordwijk: ESTEC), p.565

\bibitem[]{}
Cernicharo J., Pardo J.R., Gonz\'alez-Alfonso E. et al., 1999b, ApJL, 
520, L131
 
\bibitem[]{}
Charnley S.B., 1997, ApJ 481, 396
 
\bibitem[]{}
Cheung A.C., Rank D.M., Townes C.H., et al., 1969, Nat 221, 626
 
\bibitem[]{}
Dartois E., d'Hendecourt L., Boulanger F., et al., 1998, A\&A 331, 651
 
\bibitem[]{}
de Graauw Th., Haser L.N., Beintema D.A., et al., 1996,
A\&A 315, L49
 
\bibitem[]{}
Downes D., Genzel R., Becklin E.E., Wynn-Williams C.G., 1981, ApJ
244, 869
 
\bibitem{}
Doty S.D., Neufeld D.A., 1997, ApJ 489, 122
 
\bibitem[]{}
Evans N.J., Lacy J.H., Carr J.S., 1991, ApJ 383, 674
 
\bibitem[]{}
Feuchtgruber H., Lutz D., Beintema D.A., et al., 1997, ApJ 487, 962

\bibitem[]{}
Gensheimer P.D., Mauersberger R., Wilson T.L., 1996, A\&A 314, 281
 
\bibitem[]{}
Genzel R., Reid M.J., Moran J.M., Downes D., 1981, ApJ 244, 884
 
\bibitem[]{}
Genzel R., Stutzki J., 1989, ARAA 27, 41
 
\bibitem[]{}
Goldsmith P.F., Langer W.D., 1999, ApJ 517, 209
 
\bibitem[]{}
Gonz\'alez-Alfonso E., Cernicharo J., van Dishoeck E.F., Wright C.M.,
Heras A., 1998, ApJ 502, L169
 
\bibitem[]{}
Harwit M., Neufeld D.A., Melnick G.J., Kaufman M.J., 1998, ApJ 497, L105
 
\bibitem[]{}
Heras A.M., 1997, in: First ISO Workshop on Analytical Spectroscopy, eds. A.M.
Heras, K. Leech, N.R. Trams \& M. Perry, ESA Publications Division SP-419, 
p. 271, Noordwijk

\bibitem[]{}
Hollenbach D.J., McKee C.F., 1979, ApJS, 41, 555
 
\bibitem[]{}
Jacq T., Walmsley C.M., Henkel C., et al., 1990, A\&A 228, 447
 
\bibitem[]{}
Kaufman M.J., Neufeld D.A., 1996, ApJ 456, 611

\bibitem[]{}
Le Bourlot J., Pineau des For\^{e}ts G., Flower D.R., 1999, MNRAS 305, 802
 
\bibitem[]{}
Liseau R., Ade P., Armand C., et al., 1996, A\&A 315, L181
 
\bibitem[]{}
Melnick G.J., Stacey G.J., Genzel R., Lugten J.B., Poglitsch A., 1990,
ApJ 348, 161
 
\bibitem[]{}
Menten K.M., Reid M.J., 1995, ApJ 445, L157

\bibitem[]{}
Meyer D.M., Jura M., Cardelli J.A., 1998, ApJ 493, 222

\bibitem[]{}
Neufeld D.A., Melnick G.J., 1987, ApJ, 322, 266
 
\bibitem[]{}
Neufeld D.A., Melnick G.J., 1991, ApJ 368, 215

\bibitem[]{}
Neufeld D.A., Lepp S., Melnick G.J., 1995, ApJS, 100, 132
 
\bibitem[]{}
Nisini B., Benedettini M., Giannini T., et al., 1999, A\&A 350, 529
 
\bibitem[]{}
Phillips T.G., Scoville N.Z., Kwan J., Huggins P.J., Wannier, P.G., 1978, 
ApJ 222, L59
 
\bibitem[]{}
Rybicki G.B., Lightman A.P., 1979, Radiative Processes in Astrophysics, Wiley

\bibitem[]{}
Saraceno P., Benedettini M., Di Giorgio A.M., et al., 1999, in: The 
Physics and Chemistry of the Interstellar Medium, the 3$^{\rm rd}$ 
Cologne-Zermatt Symposium, eds. V. Ossenkopf et al., GCA-Verlag Herdecke,
p. 279

\bibitem[]{}
Schaeidt S.G., Morris P.W., Salama A., et al., 1996, A\&A 315, L55
 
\bibitem[]{}
Scoville N.Z., Hall D.N.B., Kleinmann S.G., Ridgway S.T., 1982, ApJ 253, 136
 
\bibitem[]{}
Scoville N.Z., Kleinmann S.G., Hall D.N.B., Ridgway S.T., 1983, ApJ 275, 201

\bibitem[]{}
Spinoglio L, Giannini T, Saraceno P., et al., 1999, in: 
The Universe as seen by ISO, eds. P.Cox \& M.F. Kessler, ESA 
Publications Division SP-427, (Noordwijk: ESTEC), p.517
 
\bibitem[]{}
Takahashi T., Hollenbach D.J., Silk J., 1983, ApJ 275, 145

\bibitem[]{}
Takahashi T., Hollenbach D.J., Silk J., 1985, ApJ 292, 192

\bibitem[]{}
Tauber J., Olofsson G., Pilbratt G., Nordh L., Frisk U., 1996, A\&A 308, 913
 
\bibitem[]{}
Timmermann R., Poglitsch A., Nikola T., Geis N., 1996, ApJ 460, L65
 
\bibitem[]{}
Valentijn E.A., Feuchtgruber H., Kester D.J.M., et al., 1996, A\&A 315, L60

\bibitem[]{}
van Dishoeck E.F.\ 1998, Faraday Disc. 109, 31
 
\bibitem[]{}
van Dishoeck E.F., Helmich F.P., 1996, A\&A 315, L177
 
\bibitem[]{}
van Dishoeck E.F., Wright C.M., Cernicharo J., et al., 1998, ApJ 502, L173
 
\bibitem[]{}
Wannier P.G., Pagani L., Kuiper T.B.H. et al., 1991, ApJ 377, 171
 
\bibitem[]{}
Waters J.W., Gustincic J.J., Kakar R.K., et al., 1980, ApJ 235, 57
 
\bibitem[]{}
Watson D.M., Genzel R., Townes C.H., Storey J.W.V. 1985, ApJ 298, 316
 
\bibitem[]{}
Werner M.W., Gatley I., Harper D.A., et al., 1976, ApJ 204, 420

\bibitem[]{}
Wright C.M., 1999, in: Astrochemistry: From Molecular Clouds to
Planetary Systems, proceedings of the IAU Symposium 197, eds. Y.C. Minh
\& E.F. van Dishoeck
 
\bibitem[]{}
Wright C.M., Drapatz S., Timmermann R., et al., 1996, A\&A 315, L301
 
\bibitem[]{}
Wright C.M., van Dishoeck E.F., Helmich F.P., et al., 1997, in: First 
ISO workshop on Analytical Spectroscopy, eds.\ A.\ Heras et al.\, 
ESA SP--419, (Noordwijk: ESTEC), p. 37
 
\bibitem[]{}
Wynn-Williams C.G., Genzel R., Becklin E.E., Downes D., 1984, ApJ 281, 172
 
\bibitem[]{}
Zmuidzinas J., Blake G.A., Carlstrom J., et al., 1995, in: Proceedings 
of the Airborne Astronomy Symposium on the Galactic Ecosystem: from Gas to
Stars to Dust, eds. M.R. Haas, J.A. Davidson \& E.F.\ Erickson
(San Francisco: ASP), p. 33
 
\end{thebibliography}
\end{document}